\documentclass[conference]{IEEEtran}
\IEEEoverridecommandlockouts
\usepackage[usestackEOL]{stackengine}
\usepackage{cite}
\usepackage{amsmath,amssymb,amsfonts}
\usepackage{graphicx}
\usepackage{textcomp}
\usepackage{nicefrac, xfrac}
\usepackage{xcolor}

\usepackage{tikz}
\usepackage[left=0.6in,right=0.6in,top=0.75in, bottom=0.95in
]{geometry}

\def\BibTeX{{\rm B\kern-.05em{\sc i\kern-.025em b}\kern-.08em
    T\kern-.1667em\lower.7ex\hbox{E}\kern-.125emX}}

\begin{document}

\title{Physical Layer Secret Key Generation with Kalman Filter Detrending \vspace{-0.3cm}}

\author{
	\IEEEauthorblockN{ Miroslav Mitev$^1$, Arsenia Chorti$^{2,1}$,  Gerhard Fettweis$^{1,3}$}
	\IEEEauthorblockA{ $^1$Barkhausen Institut, 01187 Dresden, Germany; \\
  $^2$ETIS, UMR 8051 CY Cergy Paris Université, ENSEA, CNRS, 95000, Cergy, France;\\ 
 $^3$Vodafone Chair for Mobile Communications Systems, Technische Universität Dresden, 01062 Dresden, Germany;\\ 
\{miroslav.mitev, gerhard.fettweis\}@barkhauseninstitut.org, arsenia.chorti@ensea.fr}
\vspace{-0.7cm}
}

\maketitle

\begin{abstract}
The massive deployment of low-end wireless Internet of things (IoT) devices opens the challenge of finding de-centralized and lightweight alternatives for secret key distribution. A possible solution, coming from the physical layer, is the secret key generation (SKG) from channel state information (CSI) during the channel's coherence time. This work acknowledges the fact that the CSI consists of deterministic (predictable) and stochastic (unpredictable) components, loosely captured through the terms large-scale and small-scale fading, respectively. Hence, keys must be generated using only the random and unpredictable part. To detrend CSI measurements from deterministic  components, a simple and lightweight approach based on Kalman filters is proposed and is evaluated using an implementation of the complete SKG protocol (including privacy amplification that is typically missing in many published works). In our study we use a massive multiple input multiple output (mMIMO) orthogonal frequency division multiplexing outdoor measured CSI dataset. The threat model assumes a passive eavesdropper in the vicinity (at 1 meter distance or less) from one of the legitimate nodes and the Kalman filter is parameterized to maximize the achievable key rate. 
\end{abstract}

\section{Introduction}
The sixth generation wireless networks (6G) are anticipated to facilitate the extensive deployment of Internet of Things (IoT) devices. However, the high computational demands of many cryptographic schemes, particularly in the field of public key encryption (PKE), can significantly impact performance and drain the battery of power-limited devices~\cite{Amitav}, \cite{Aylin}. Furthermore, the emergence of quantum computing makes existing PKE algorithms insecure. To address this concern, physical layer security (PLS)-based secret key generation (SKG) is identified as a viable quantum-secure alternative. This lightweight approach (first proposed in \cite{Maurer} and \cite{Csiszar}) allows the extraction of shared randomness directly from the wireless channel and provides means for secure communication.

Despite the existence of vast theory behind the SKG scheme, real-world implementations are scarce. The openness of the wireless medium creates a challenge, as eavesdroppers in the vicinity might observe similar channel state information (CSI) and obtain partial knowledge on the ``secret'' key if information leakage is not explicitly considered. Therefore, correlations between legitimate and adversarial channel observations must be taken into account. To  ensure independence in time, frequency and antenna domains, simple approaches, e.g., subsampling, can be used. However, accounting for dependencies in space and in particular near-by locations, requires explicit consideration. 

While small-scale fading effects de-correlate rapidly over short distances, large-scale fading phenomena vary slowly and can remain stable in time~\cite{Andrea_Goldsmith_book} (making it predictable by nearby nodes~\cite{RSS_prediction}). In this sense, it is important that large-scale effects are removed from the CSI and keys are generated only from the unpredictable and random part~\cite{Context_aware_security_Chorti2022}. Recent work has presented pre-processing steps to address this issue~\cite{Srinivisan21}. The authors propose the use of principal component analysis (PCA) and autoencoders (AE) to separate the components of the channel. However, such algorithms might require high computational power~\cite{PCA_complexity}, making them unsuitable for low-end devices.

In this work, we present a lightweight detrending approach based on Kalman filters. Kalman filter is a standard technique to smooth noisy measurements and extract location-dependent trends. Such trends are mainly represented by path-loss and shadowing (i.e., large-scale fading). Following from that, we propose to isolate the entropy rich, small-scale fading, present in the wireless channel, by treating the output of the Kalman filter as a predictable component which must be removed  before SKG. This concept was first introduced in our earlier work~\cite{Mitev2022_access}, where SKG was combined with location information in a zero round-trip-time (0-RTT) authentication protocol. In~\cite{Mitev2022_access} we focused mainly on authentication and provided security proofs assuming  that secret keys can be generated at sufficient rates.  

The current work is a proof of concept aiming at demonstrating the feasibility of executing a lightweight SKG. All steps of the SKG protocol are implemented and evaluated using massive multiple input multiple output (mMIMO) real-life outdoor measurements provided by Nokia Bell-Labs~\cite{Nokia_data}. The dataset consists of CSI measurements of mobile users that pass by nearby locations that are 1 meter (or less) apart but are separated in time. Each user generates secret keys from CSI measurements which are reconciled using Slepian-Wolf implementations of Polar codes. This work aims at evaluating spatial correlations in time, i.e., we evaluate leakages to malicious users who pass by similar location as legitimate users but at different time instances. To derive the final SKG rate we perform conditional min-entropy evaluation on the legitimate reconciled sequences with respect to the information obtained by the attacker in order to determine the required amount of privacy amplification.

The rest of this paper is organized as follows: Section \ref{sec:sys_model} gives a detailed overview of the SKG protocol. It also presents our system model and introduces the proposed Kalman filter-based randomness extraction. Section \ref{sec:dataset}, presents the details on the dataset used for this study. Next, a step-by-step evaluation of the SKG protocol is given in Section \ref{sec:results}. Finally, Section \ref{sec:conclusion} concludes this paper.

\section{Kalman Filter Based Deternding for Secret Key Generation} \label{sec:sys_model}

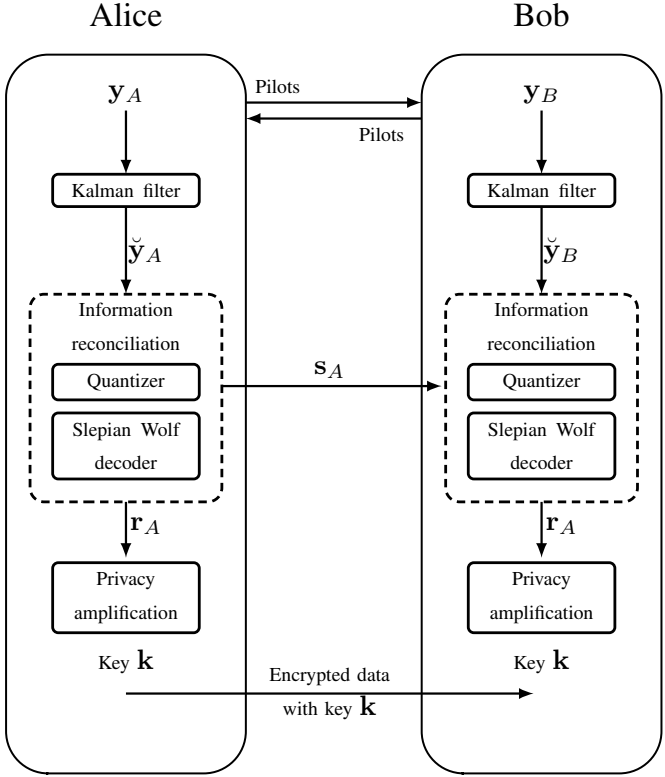
\begin{figure}[t]
\centering
\tikzset{>=latex}
 \resizebox{9cm}{!}{
\begin{tikzpicture}
\begin{scope}[xshift=2cm]
\draw[rounded corners=15pt,line width=0.71pt] (0.5,0)--(3,0)--(3,9)--(0,9)--(0,0)--(.5,0);
\node at (1.5,9.5) {\large Alice};
\draw[thick, ->] (3,8.4) -- (5.2,8.4);
\draw[thick, ->] (5.2,8.2)--(3,8.2);
\node at (3.4,8.6) {\scriptsize Pilots};
\node at (4.7,8) {\scriptsize Pilots};
\node at (1.5,8.5) {$\mathbf{y}_A$};
\draw[thick, ->] (1.5,8.3)--(1.5,7.5);

\node[draw, rounded corners=2pt,line width=1pt, text width=1.6cm, align=center] at (1.5,7.3) {\Longstack[c]{\scriptsize Kalman filter}};
\draw[thick, ->] (1.5,7.1)--(1.5,6);
\node at (1.75,6.55) {$\breve{\mathbf{y}}_A$};
\node[draw, rounded corners=2pt,line width=1pt, text width=1.6cm, align=center] at (1.5,4.9) {\Longstack[c]{\scriptsize Quantizer}};

\node[draw, rounded corners=2pt,line width=1pt, text width=1.6cm, align=center] at (1.5,4.1) {\Longstack[c]{\scriptsize Slepian Wolf \\ \scriptsize decoder }};

\draw[rounded corners=5pt,line width=1pt,densely dashed] (1,3.4)--(2.7,3.4)--(2.7,6)--(0.3,6)--(0.3,3.4)--(1,3.4);

\node at (1.5,5.6) {\Longstack[c]{\scriptsize Information \\ \scriptsize reconciliation}};

\draw[thick, ->] (2.7,4.85)--(5.45,4.85);
\node at (4.05,5.05) {$\mathbf{s}_A$};

\draw[thick, ->] (1.5,3.4)--(1.5,2.7);
\node at (1.75,3.1) {$\mathbf{r}_A$};
\node[draw, rounded corners=2pt,line width=1pt, text width=1.6cm, align=center] at (1.5,2.2) {{\scriptsize Privacy \\ \scriptsize amplification }};
\node at (1.5,1.4) {\scriptsize Key \normalsize  $\mathbf{k}$};
\draw[thick, ->] (1.5,1)--(6.6,1);
\node at (4.05,1) {\Longstack[c]{\scriptsize Encrypted data  \\ \scriptsize with key \normalsize $\mathbf{k}$}};
\end{scope}

\begin{scope}[xshift=7.2cm]
\draw[rounded corners=15pt,line width=0.71pt] (0.5,0)--(3,0)--(3,9)--(0,9)--(0,0)--(.5,0);
\node at (1.5,9.5) {\large Bob};
\node at (1.5,8.5) {$\mathbf{y}_B$};
\draw[thick, ->] (1.5,8.3)--(1.5,7.5);

\node[draw, rounded corners=2pt,line width=1pt, text width=1.6cm, align=center] at (1.5,7.3) {\Longstack[c]{\scriptsize Kalman filter}};
\draw[thick, ->] (1.5,7.1)--(1.5,6);
\node at (1.75,6.55) {$\breve{\mathbf{y}}_B$};
\node[draw, rounded corners=2pt,line width=1pt, text width=1.6cm, align=center] at (1.5,4.9) {\Longstack[c]{\scriptsize Quantizer}};

\node[draw, rounded corners=2pt,line width=1pt, text width=1.6cm, align=center] at (1.5,4.1) {\Longstack[c]{\scriptsize Slepian Wolf \\ \scriptsize decoder }};

\draw[rounded corners=5pt,line width=1pt,densely dashed] (1,3.4)--(2.7,3.4)--(2.7,6)--(0.3,6)--(0.3,3.4)--(1,3.4);

\node at (1.5,5.6) {\Longstack[c]{\scriptsize Information \\ \scriptsize reconciliation}};
\draw[thick, ->] (1.5,3.4)--(1.5,2.7);
\node at (1.75,3.1) {$\mathbf{r}_A$};
\node[draw, rounded corners=2pt,line width=1pt, text width=1.6cm, align=center] at (1.5,2.2) {{\scriptsize Privacy \\ \scriptsize amplification }};
\node at (1.5,1.4) {\scriptsize Key \normalsize  $\mathbf{k}$};
\end{scope}
\end{tikzpicture}
}
\caption{Secret key generation between Alice and Bob using Kalman filter. Thanks to reciprocity in wireless channels, the outputs of the quantizers are highly correlated variables but not the same (due to noise). Using $\mathbf{s}_A$ and his Slepian Wolf decoder Bob corrects the errors to obtain $\mathbf{r}_{A}$. Final keys are obtained after privacy amplification.} \label{fig:Miro SKG}
\end{figure}

The system model in this work consists of two legitimate users, Alice and Bob and a malicious user within the network, Eve. A sketch of the SKG protocol used in this paper is depicted in Figure \ref{fig:Miro SKG} and described below.
\par\noindent1) \textit{Advantage distillation with Kalman filtering:} To estimate their reciprocal CSI Alice and Bob exchange orthogonal frequency-division multiplexing (OFDM) probe signals in a time-division duplex manner\footnote{In multiple attack scenarios it has been demonstrated that Alice and Bob should optimally use equal power distribution for channel probing~\cite{Mitev_Entropy2021}.}. Due to noise and possible imperfect CSI estimation, the two estimates will be different. At different point in time, Eve passes in a similar location as Bob and aims at obtaining correlated measurements. The complex signals received at Alice, Bob and Eve can be denoted as: 
\begin{align}
    \mathbf{y}_A = \mathbf{h} X + \mathbf{n}_A,\\
    \mathbf{y}_B = \mathbf{h} X + \mathbf{n}_B,\\
    \mathbf{y}_E = \mathbf{h}_E X + \mathbf{n}_E,
\end{align}
where $X \in \mathbb{C}$ is the transmit probe symbol, $\mathbf{n}_A, \mathbf{n}_B, \mathbf{n}_E \in \mathbb{C}^{N \times 1}$ are additive white Gaussian noise variables and $\mathbf{h}, \mathbf{h}_E \in \mathbb{C}$ denote the channel coefficients between Alice and Bob, Alice and Eve, respectively.
We note that real-world measurements were used to produce the results in this work (as opposed to simulated channel models), hence, no assumptions on the PDFs of the variables above can be made.

To extract randomness from the channel we propose a lightweight fast Kalman filter-based approach. Fast Kalman filter has computational complexity of $\mathcal{O}(N)$~\cite{ fast_kalman_complexity}, which allows for real-time execution on resource constrained devices~\cite{ real-time_Kalman_book}. The filter assumes that a  state $G_{A,m}$ is related to the previous $G_{A,m-1}$ as:  
\begin{equation}
    G_{A,m}=G_{A,m-1}+K_{A,m}(Y_{A,m}-G_{A,m-1}),
    \label{eq:Kalman}
\end{equation}
where $Y_{A,m}$, $G_{A,m}$, for sample index $m=1,2,\dots, M$ are the values of raw and filtered  measurements, respectively, and $K_{A,m}$ is Kalman gain which determines the convergence of the filter. The Kalman gain is computed as:
\begin{equation}
K_{A,m}=\frac{P_{A,m}}{P_{A,m}+{R}}, \label{eq:k_gain} 
\end{equation}
where $P_{A,m}$ is a prediction error which  updates iteratively during the filtering process and $R$ denotes variance of the expected error in the raw measurement data, this is a pre-defined constant. From \eqref{eq:Kalman} and \eqref{eq:k_gain} it is clear that $R$ has an important role and defines how much to ``trust'' the raw measurements. At the end of this step Alice obtains a vector $\mathbf{g}_A=[G_{A,1}, \dots, G_{A,M}]$ containing the filter output which is of the same size as her raw measurement vector $\mathbf{y}_A=[Y_{A,1}, \dots, Y_{A,M}]$. To remove predictable components, Alice subtracts the filter output from her raw measurements and obtains the residual $\breve{\mathbf{y}}_A=\mathbf{y}_A-\mathbf{g}_A$. The process is defined identically for Eve and Bob who obtain $\breve{\mathbf{y}}_E$ and $\breve{\mathbf{y}}_B$, respectively.

\begin{figure}[!t]
    \centering
    \includegraphics[width=0.48\textwidth]{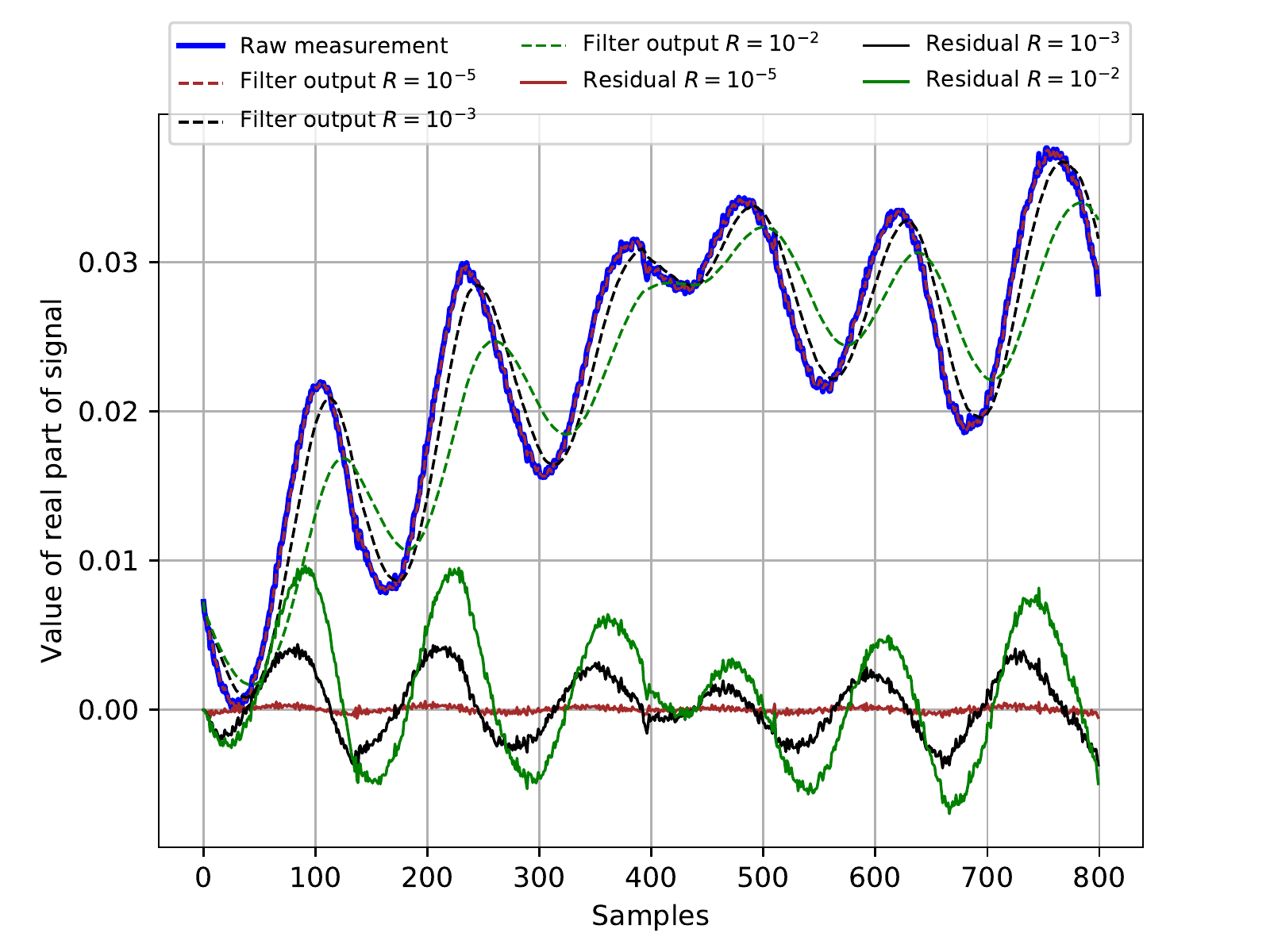}
    \caption{Kalman filtering outputs for $R=10^{-2}$, $R=10^{-3}$, $R=10^{-5}$. The resulting residuals (i.e., subtracting raw and filtered data) are also illustrated. }
    \label{fig:Kalman}
\end{figure}

To demonstrate how the value of $R$ affects the filter output, Figure \ref{fig:Kalman} shows an example for $R=10^{-2}$, $R=10^{-3}$  and $R=10^{-5}$. The figure illustrates raw measurements, filter outputs and residuals (that are result of subtracting the previous two vectors).
\begin{figure*}
    \centering
    \includegraphics[ width=0.85\textwidth]
        {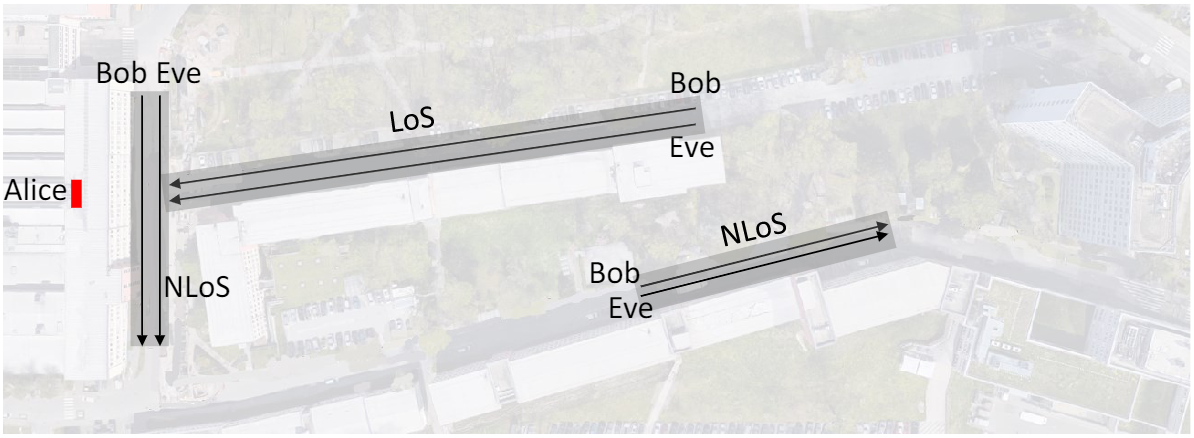}
    \caption{Snapshot of the measurement campaign from~\cite{Nokia_data}. Only the highlighted tracks are considered in this work.}
    \label{fig:NOKIA_campus}
\end{figure*}

It can be seen that depending on the value of $R$ the filter output can follow the raw measurements loosely (for $R=10^{-2}$) or closely (for $R=10^{-5}$) resulting into residual with large or small-scale variations, respectively. It can be seen that when $R=10^{-2}$ the residual follows the trend of the original data. Next, when $R=10^{-3}$ the residual becomes closer to zero-mean and large predictable variations are minimized. Finally, for $R=10^{-5}$, the residual is almost constant with values close to zero. This shows that a large value of $R$ might result in leaving predictable components in the residual, however, if $R$ becomes too small the filtering removes not only predictable but also random components from the raw measurements.

\par\noindent2) \textit{Information reconciliation:} Alice, Bob and Eve quantize their observations to binary vectors $\mathbf{r}_A, \mathbf{r}_B, \mathbf{r}_E$, respectively. To correct errors at the output of the quantizer one of the legitimate users (Alice)  generates syndrome information, $\mathbf{s}_A$, using distributed source coding techniques (e.g., Slepian-Wolf coding). The syndrome is send to the other legitimate party (Bob) on a public channel. Bob uses the syndrome to correct errors in his observations using a DSC decoder. Considering successful reconciliation, at the end of the step Alice and Bob possess identical sequence, $\mathbf{r}_A$. Due to the public transmission we assume that $\mathbf{s}_A$ is also fully accessible to Eve. Using the syndrome she tries to correct errors in her observations, $\mathbf{r}_E$. At the output of her decoder, Eve obtains $\mathbf{r}_E'$ which, depending on initial channel correlations, could be close or not to $\mathbf{r}_A$.

\par\noindent3) \textit{Privacy amplification:} This step is performed to remove leakage that occurred in the previous steps. The length of the final key $\mathbf{k} \in \mathcal{K}$ between Alice and Bob should be~\cite{min_entropy, how_much_hash}:
\begin{equation}
    |\mathbf{k}| \leq H_{\infty}(\mathbf{r}_A|\mathbf{r}_E, \mathbf{s}_A, \mathbf{r}_E'), \label{eq:key_size}
\end{equation}
where~\cite{cme}: 
\begin{equation}
    H_{\infty}(\mathbf{r}_A|\mathbf{r}_E, \mathbf{s}_A, \mathbf{r}_E')\!=\!-\!\log_2\!\!\!\!\!\max_{\mathbf{r}_A, \mathbf{r}_E, \mathbf{r}_E' \in \mathcal{R}, \mathbf{s}_A \in \mathcal{S}}\!\!\!\!\!\!p(\mathbf{r}_A|\mathbf{r}_E, \mathbf{s}_A, \mathbf{r}_E'), \label{eq:cme}
\end{equation}
denotes conditional min-entropy, and $\mathcal{R}, \mathcal{S}$ denote the space of quantization outputs and syndromes, respectively. The total amount of leaked information to Eve can be evaluated as~\cite{Quant_inf_flow}:
\begin{equation}
    \text{Leakage} = H_{\infty}(\mathbf{r}_A) - H_{\infty}(\mathbf{r}_A|\mathbf{r}_E, \mathbf{s}_A,\mathbf{r}_E'),
\end{equation}
where $H_{\infty}(\mathbf{r}_A)$ is min-entropy of the sequence $\mathbf{r}_A$. A standard way to remove the leakage from the reconciled information is by using a  one-way collision-resistant compression function, e.g., hash function. This last phase ensures that the generated key sequence is uniformly distributed and unpredictable by an adversary~\cite{Mitev_EURASIP2020}.

\section{Experimental Campaign and Dataset Description} \label{sec:dataset}

The dataset used to conduct this study comes from a measurement campaign done by Nokia crew on their campus in Stuttgart, Germany~\cite{Nokia_data}. It consists of outdoor mMIMO channel measurements taken  while moving along different paths. The paths we consider in this study are illustrated in Figure \ref{fig:NOKIA_campus}. The transmit antenna array (Alice) is placed on a roof top of a building. The array has $64$ antenna elements with a rectangular geometry of $4$ rows each with $16$ single-polarization patch antennas. Horizontal and vertical spacing between antenna elements are $\lambda/2$ and $\lambda$, respectively.

In this setting, Alice transmits $64$ pilot signals following the $10$~MHz LTE numerology ($600$ subcarriers with $15$~kHz spacing). The waveform is OFDM with center frequency of $2.18$~GHz. To obtain channel measurements $50$ sub-bands are sounded, i.e., $12$ consecutive subcarriers are used per sub-band. A pilot burst over all sub-bands requires $0.5$~ms, hence bursts are sent with periodicity of $0.5$~ms. This results in $1$ ms to perform two-way exchange (e.g., Alice - Bob and Bob - Alice). 

Two user equipment (Bob and Eve) are mounted on mobile carts, each having a single monopole antenna at $1.5$~m. Users are equipped with Rohde \& Schwarz TSMW receivers and Rohde \& Schwarz IQR hard disc recorders which continuously measure and store the signals from Alice. As depicted in Figure \ref{fig:NOKIA_campus}, Bob and Eve move in parallel tracks; note the distance between Bob's and Eve's tracks is kept $\leq 1$~m. During measurements, devices are synchronized via GNSS. Bob and Eve move along the tracks at different time instances but at identical  speed of $3.6$~km/h. In accordance to the periodicity of the pilot bursts this results in approximately $0.1$~mm of sampling in the spacial domain.

In a previous work it has been demonstrated that by subsampling in frequency, time and antenna domains correlations can be removed~\cite{TMLCN}. Similarly here, to account for correlation along these domains we consider measurements at every $4$-th antenna, at every $10$-th subcarrier, and we keep every $5$-th channel sample. Taking every $5$-th samples results in time sampling factor $T=5$~ms. Furthermore, as the dataset contains only uplink measurements we use subsequent samples to mimic the downlink, i.e., odd samples are considered as downlink and even samples are uplink. A statistical analysis on the dataset can be found in~\cite{TMLCN}.

At this point, each party possesses a vector of raw channel measurements over which applies fast Kalman filter detrending with parameter $R$. As discussed in Section \ref{sec:sys_model}, Alice, Bob and Eve then subtract the output of the filter from their initial raw channel measurements and obtain the residual vectors  $\breve{\mathbf{y}}_A, \breve{\mathbf{y}}_B$ and $ \breve{\mathbf{y}}_E$, respectively. To perform quantization and reconciliation and arrive at the desired size, the vectors at each party are reshaped into a matrix of size $\frac{|\breve{\mathbf{y}}_A|}{512} \times 512$. Rows from the resulting matrices are quantized independently. In this work, we assume a linear quantizer with $4$ quantization levels. This gives $\log_2(4)=2$ bits per sample, hence, each row of the matrix produces a sequence of $1024$ bits. The quantization levels are chosen uniformly between the minimum and maximum values in the corresponding row. The resulting matrices of quantized residuals are used for key generation. In the next section we evaluate each step of the SKG protocol and show how the proposed Kalman filtering approach affects the performance.

\section{Performance Evaluation} \label{sec:results}
\begin{table}[!t]
    \caption{Mismatch probability after quantization at all parties considering different values for $R$.}
    \centering
    \begin{tabular}{|l|c|c|c|c|}
    \hline 
    Nodes &\multicolumn{2}{|c|}{Alice and Bob} &\multicolumn{2}{|c|}{Eve}  \\ \hline
    Filtering parameter& LoS & NLoS & LoS & NLoS \\ \hline
    No filtering & $0.019$ & $0.037$ & $0.51$ & $0.47$ \\ \hline
    $R=10^{-1}$  & $0.023$ & $0.039$ &  $0.47$ & $0.46$ \\ \hline
    $R=10^{-2}$  & $0.030$ & $0.046$ &  $0.46$ & $0.45$\\ \hline
    $R=10^{-3}$  & $0.040$ & $0.062$ &  $0.47$ & $0.45$\\ \hline
    $R=10^{-4}$  & $0.054$ & $0.091$ &  $0.48$ & $0.45$\\ \hline
    $R=10^{-5}$  & $0.072$ & $0.124$ &  $0.49$ & $0.47$\\ \hline
    $R=10^{-6}$  & $0.083$ & $0.143$ &  $0.49$ & $0.47$\\ \hline
    \end{tabular}
    \label{tab:mismatch}
\end{table}
Following from the previous section, we first evaluate the the mismatch probability between Alice and Bob. After quantizing mismatch is measured using Hamming distance over all generated bits. The resulting probability values are given in Table \ref{tab:mismatch}. As expected decreasing $R$ increases the mismatch between Alice and Bob. A smaller value of $R$ can make the residual more unpredictable, such that it does not follow the trend of the original measurements (see Figure \ref{fig:Kalman}), however, as observed here, this comes at the cost of decreased reciprocity. This can be observed for both LoS and NLoS conditions. On the other hand, the mismatch at Eve remains almost stable around $50\%$. The filtering process has a negligible impact on her mismatch probability. This is a desired behavior as the Kalman filter does not bring improvement at her end. 

\begin{figure}[!t]
\centering
\begin{tikzpicture}
\node[inner sep=0pt] at (0,4.5)
{\includegraphics[clip, trim=0.4cm 5.9cm 0cm 0cm,width=0.44\textwidth]{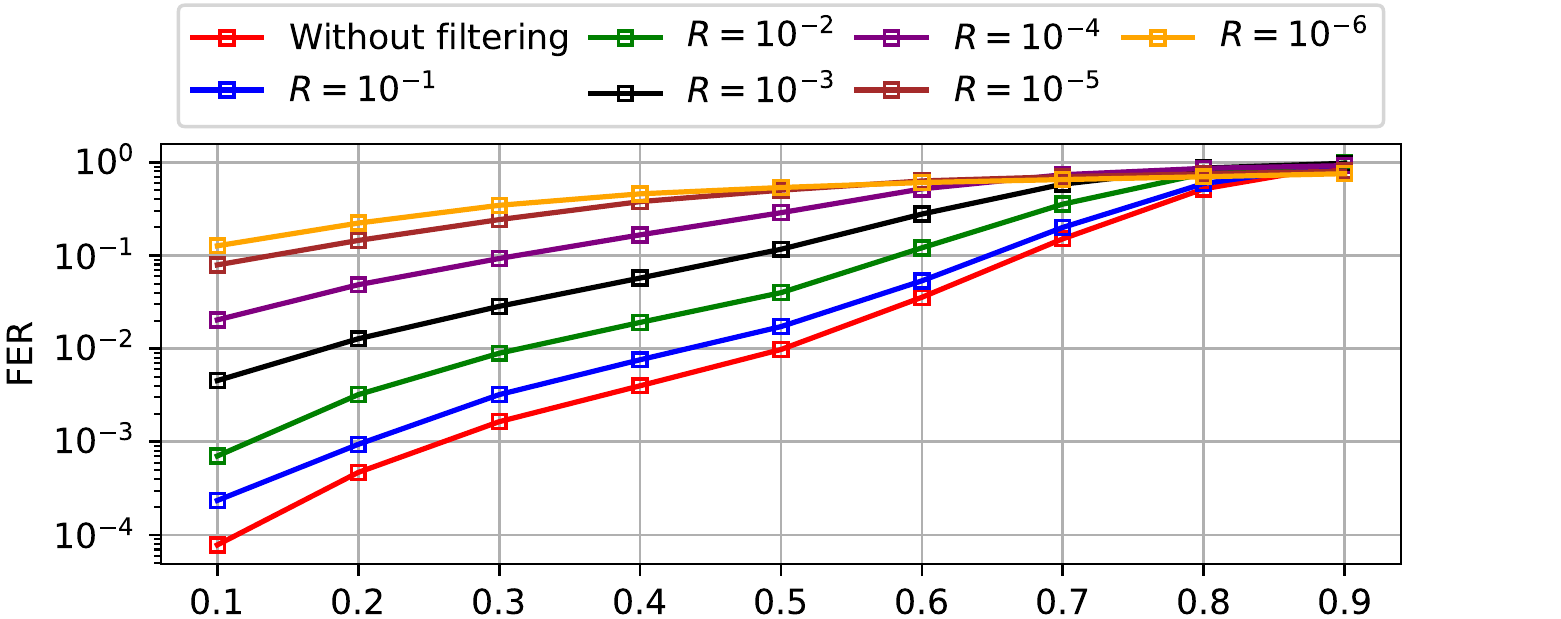}};
\node[inner sep=0pt] at (0,0.4)
{\includegraphics[clip, trim=0.4cm 0cm 0cm 6.5cm,width=0.44\textwidth]{Figures/Fff_NLOS1024_4.pdf}};
\node at (0.2,2.4) {{\footnotesize a) Frame error rate in LoS conditions}};
\node at (0.2,-1.4) {{\footnotesize b) Frame error rate in NLoS conditions}};
\end{tikzpicture}
\caption{Frame error rate in LoS and NLoS conditions after reconciliation between Alice and Bob. Different code rates and filtering parameters are illustrated.}
\label{fig:FER}
\end{figure}
After quantization the parties perform information reconciliation using Slepian Wolf implementation of Polar codes. Particularly, Polar codes with unique decoding as in~\cite{Mahdi_reconciliation_codes21} are used. Here, Alice generates a syndrome $\mathbf{s}_A$ and sends it to Bob. The syndrome size varies depending on the code rate, i.e., low code rates require longer syndrome. This gives better chances for successful reconciliation, however, leaks more information as $\mathbf{s}_A$ is also observed by Eve. At this step we evaluate the frame error rate (FER) between Alice and Bob.

This is ilustrated in Figure \ref{fig:FER}. The figure shows the FER for both LoS (Figure \ref{fig:FER}a) and NLoS (Figure \ref{fig:FER}b) scenarios. It can be seen that the FER decreases in NLoS conditions. This is an expected result, as the absence of the dominant LoS path naturally decreases the SNR between the two parties. As FER is directly impacted by the bit mismatch probability, we see a similar behaviour as in Table \ref{tab:mismatch}, i.e., a smaller value of $R$ gives a lower performance in terms of FER. Another important parameter for the success of this step is code rate. At lower code rate the FER is negligible, however, as noted above this requires the exchange of larger syndrome sequence $\mathbf{s}_A$.
 
\begin{figure}[!t]
    \centering
    \includegraphics[width=0.48\textwidth]{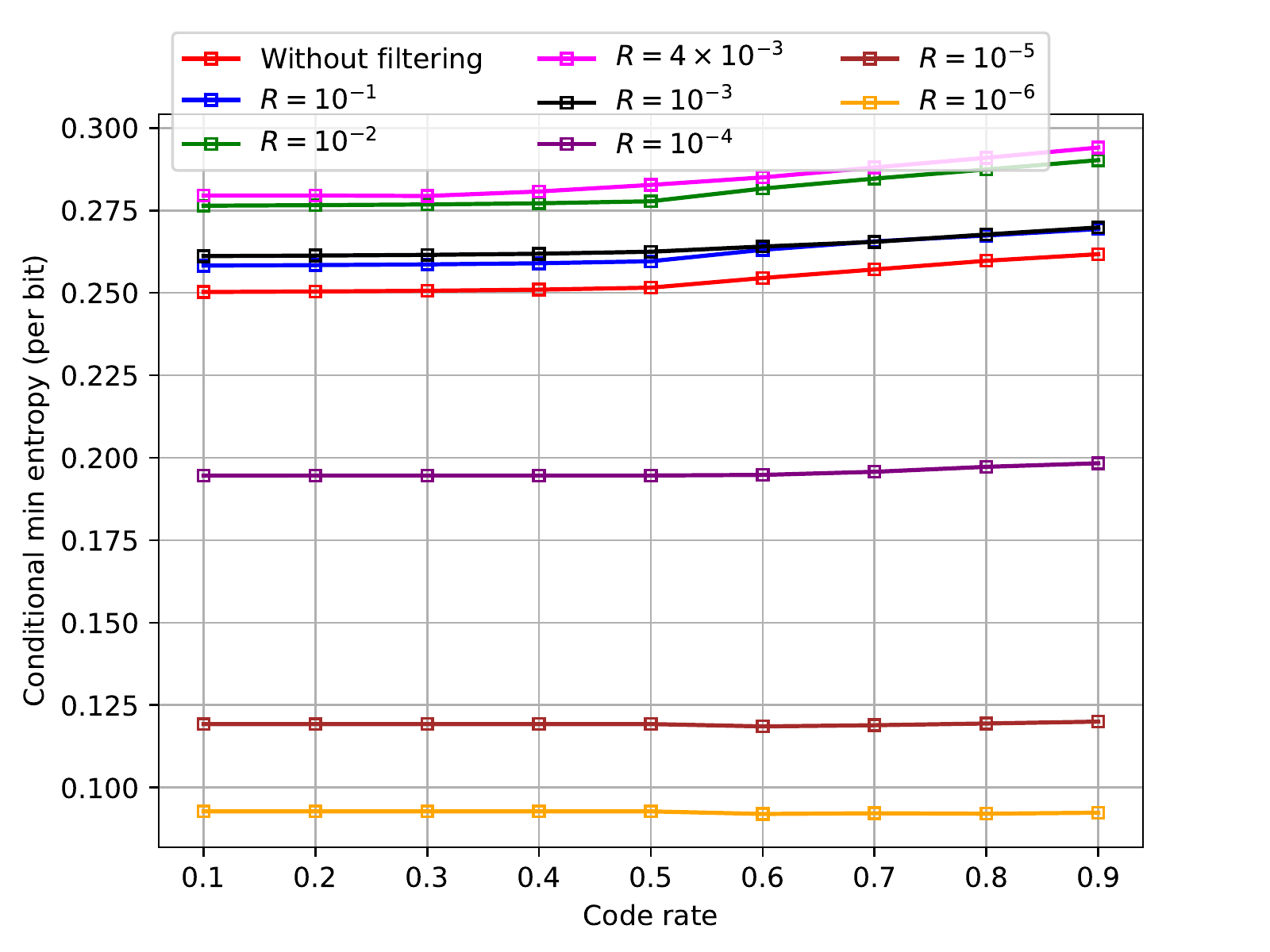}
    \caption{Conditional min-entropy in LoS condition considering different code rates and values of $R$.}
    \label{fig:CME_LOS}
\end{figure}

\begin{figure}[!t]
    \centering
    \includegraphics[width=0.48\textwidth]{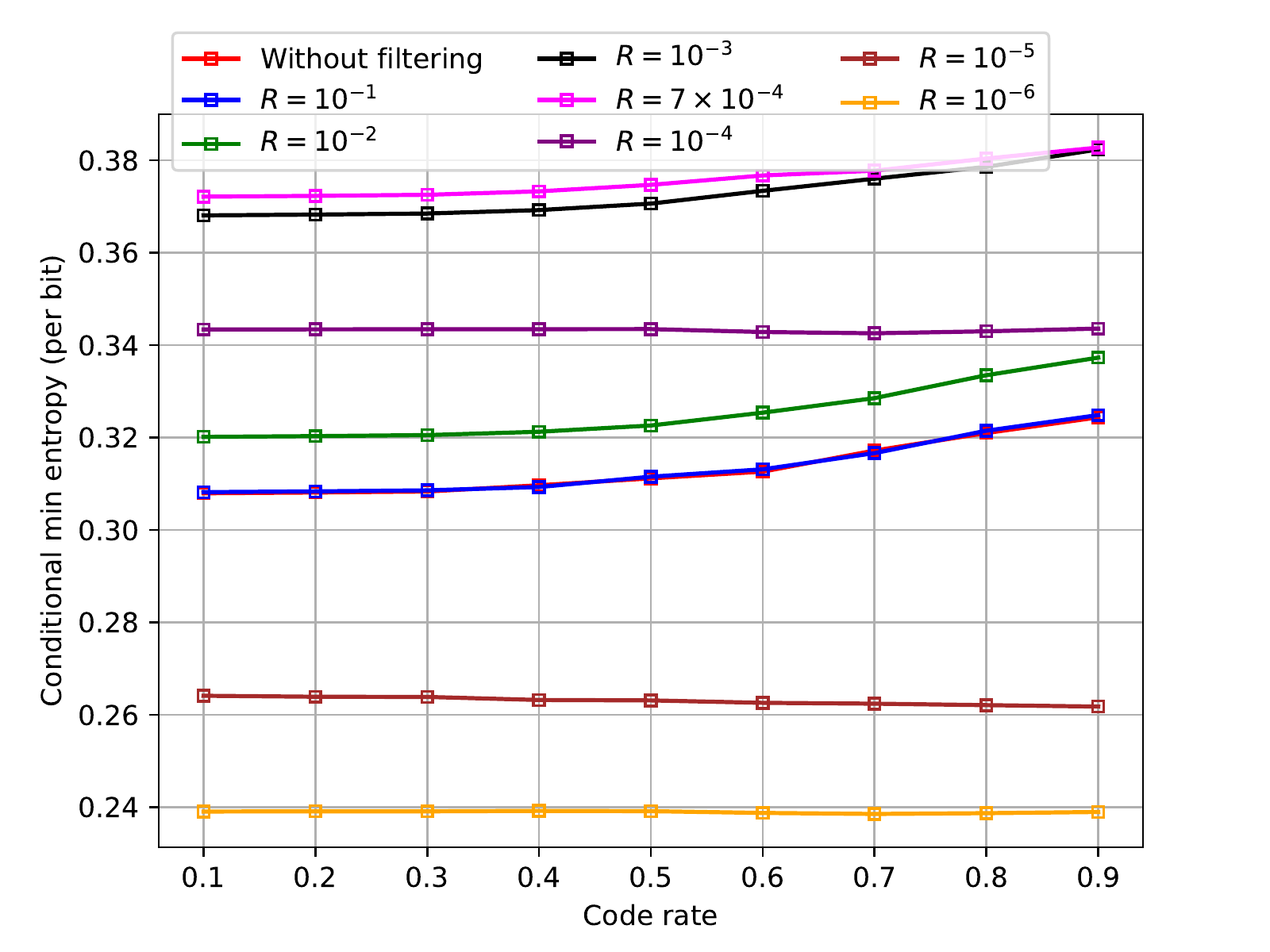}
    \caption{Conditional min-entropy in NLoS condition considering different code rates and values of $R$.}
    \label{fig:CME_NLOS}
\end{figure}

After reconciliation we evaluate the secure and random number of bits by considering the leakage at Eve. In our work, this is performed using the FBLEAU estimator~\cite{FBLEAU}. The evaluation is done in accordance to Equation~\eqref{eq:cme}, i.e., the estimator takes as inputs $\mathbf{r}_A, \mathbf{r}_E, \mathbf{s}_A, \mathbf{r}_E'$ and outputs a scalar value for the conditional min-entropy. The results are illustrated in Figure \ref{fig:CME_LOS} for LoS and Figure \ref{fig:CME_NLOS} for NLoS. Several observations can be drawn from the figures.

First, we can see that LoS can in general offer less randomness, hence, lower conditional min entropy. This may be attributed to the fact that main contributors for unpredictability in wireless channels are multipath components (MPCs) that arrive at different times and result from different reflectors. While MPCs are present also in the LoS setup, the received signals are mainly affected by direct path as she contains the most of the power. On the other hand, in NLoS conditions the reflected components are more pronounced which results in higher unpredictability of the received signals.

Second, applying Kalman filter can improve the performance in terms of conditional min-entropy. However, this is valid only up to a certain $R$ value. For the LoS scenario we see that applying the filter with $R=10{-1}$ gives a slight improvement. The increase in conditional min-entropy continuous for $R=10{-2}$ and $R=4 \times 10^{-3}$ (this value was found with numerical search in the space). However, after these values a sudden drop  is observed reaching conditional min-entropy $<0.1$ for $R=10^{-6}$. A similar tren is observed for the NLoS scenario shown in Figure~\ref{fig:CME_NLOS} with the optimal value identified, $R=6 \times 10^{-4}$. Given these results it is clear that an optimal value of $R$ exists and it is determined by the channel conditions. In the LoS case less filtering is required to separate the unpredictable components as compared to the NLoS case. Similar, to the conclusion above we believe that this is a result of the lower randomness in LoS scenarios. 

Finally, we can see that the code rate could also affect the conditional min-entropy. This result is expected as increasing the code rate decreases the length of the syndrome. As noted earlier, the syndrome is shared on a public channel and is available to Eve. Therefore, higher code rate leaks less information and brings an increase in  conditional min-entropy.

\begin{figure}[!t]
    \centering
    \includegraphics[width=0.48\textwidth]{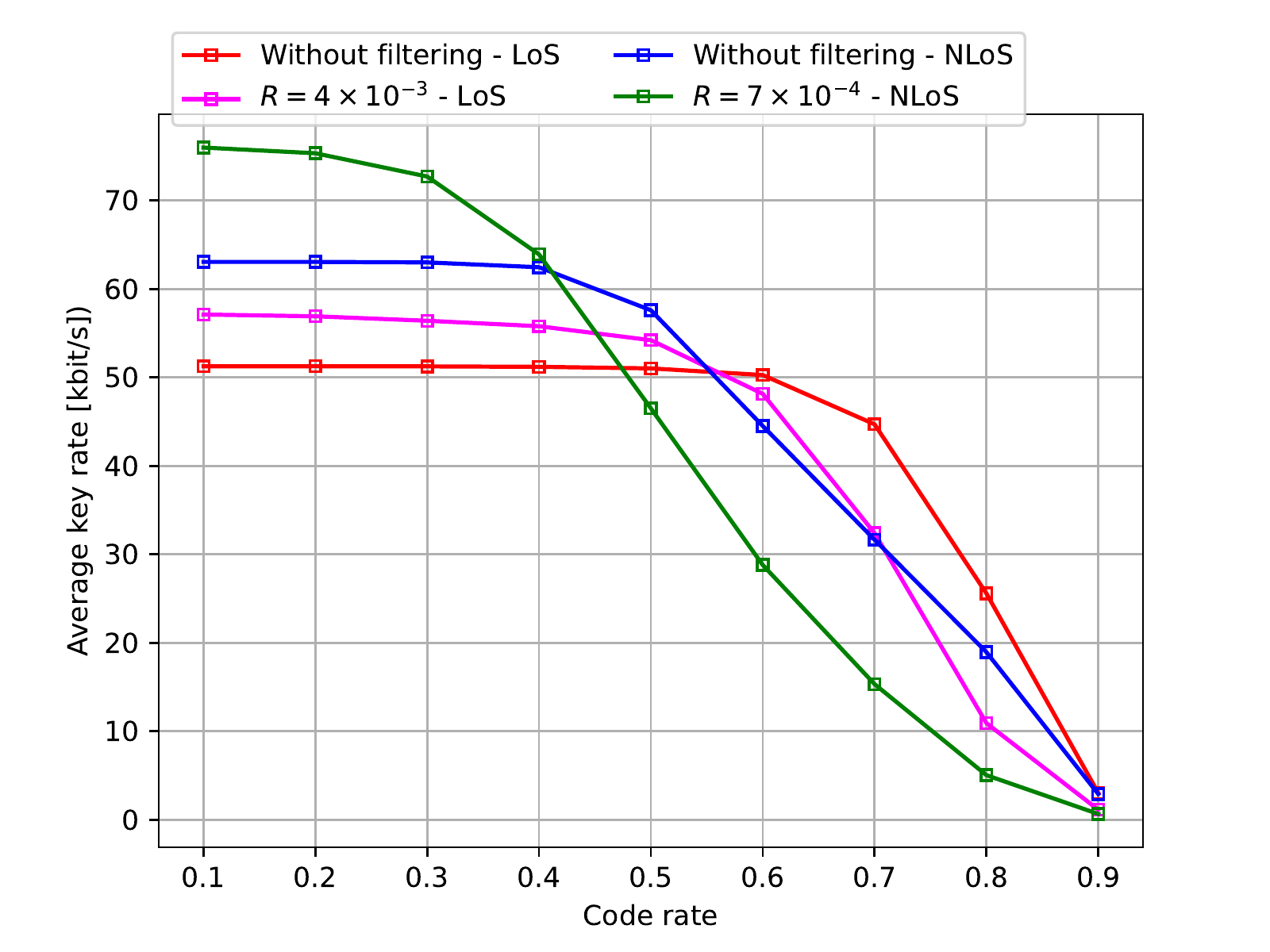}
    \caption{Average key rate for LoS and NLoS conditions considering different code rates and filtering paratemers.}
    \label{fig:key_rate}
\end{figure}
At this step we can evaluate the average key rate for the presented scenarios. For the setup in this work we represent the key rate in $[b/s]$ as a function of several parameters: i) number of bits generated at the output of the quantizer $=1024$; ii) FER after performing  reconciliation; iii) conditional min-entropy as denoted in Equation \eqref{eq:cme}; iv) time sampling factor, $T$, which as noted in Section \ref{sec:dataset} for the current setup equals to $5$ ms. Based on that the average key rate is defined as:
\begin{equation}
    R [b/s] = \frac{F \times (1-\text{FER}) \times H_{\infty}(\mathbf{r}_A|\mathbf{r}_E, \mathbf{s}_A, \mathbf{r}'_E)}{T}.
\end{equation}
Figure \ref{fig:key_rate} provides an evaluation using subset of the combinations above. The figure shows the average key rate in LoS and NLoS conditions when Kalman filter is applied (with parameters $R=4 \times 10^{-3}$ and $R=7 \times 10^{-4}$) and when it is not. The values of $R$ are chosen in accordance to the results on conditional min-entropy. For low code rates it can be seen that the filter  provides substantial improvement. Particularly, improvement of $>5$ [kbit/s] for LoS and $>10$ [kbit/s] for NLoS. It can also be observed that increasing the code rate results in decrease in performance. This aligns with our results in Figure \ref{fig:FER} as high code rate corresponds to high FER. 

Finally, as noted in Section \ref{sec:sys_model}, to generate the final keys, hashing must be performed. We execute this step using SHA-256 hashing function. The output of SHA-256 has a fixed length of $256$ bits. To comply with Equation \ref{eq:key_size} we fix the input of the function to size $256/H_{\infty}(\mathbf{r}_A|\mathbf{r}_E, \mathbf{s}_A, \mathbf{r}'_E)$. After hashing we verify the randomness of the generated keys, by passing them through the NIST randomness test collection~\cite{NIST_randomness}. The tests  evaluate uniformity, independence and unpredictability, of the key bits and output a binary decision (yes/no). All generated keys for LoS $R=4 \times 10^{-3}$ and NLoS $R=7 \times 10^{-4}$ and code rate $0.1$ are evaluated. The resulting success rates are given in Table~\ref{tab:NIST}. We can see that the values are close to one, proving that the generated keys are high randomness properties.

\begin{table}[!t]
    \caption{Randomness evaluation using NIST-approved tests~\cite{NIST_randomness}.}
    \centering
    \begin{tabular}{|l|c|}
    \hline
         Test & Success rate    \\
         \hline
         Frequency (monobit) test & 0.9926   \\
         \hline
         Frequency within a block test & 0.9838   \\
         \hline
         Runs test & 0.9868  \\
         \hline
         Longest run of ones in a block test & 0.9868  \\
         \hline
         Serial test & 0.9874  \\
         \hline
         Cumulative sum test & 0.9926  \\
         \hline
    \end{tabular}
    \label{tab:NIST}
\end{table}

Overall, our results demonstrate that Kalman filter can be an efficient and lightweight approach to extract random components from the wireless channel. As a future work we plan to further investigate our approach by optimizing the parameterization throughout the SKG protocol.

\section{Conclusion} \label{sec:conclusion}
This work provides an experimental validation of the PLS-based secret key generation. All steps of the protocol are performed on a real-life outdoor dataset. In general, our findings illustrate that the utilization of the Kalman filter can be a viable method for extracting random elements from the wireless channel. Based on our evaluation, it is evident that the information accessible for SKG is highly dependent on the characteristics of the channel, i.e., in LoS or NLoS. Therefore devices need to be channel-aware and their system parameters must be chosen accordingly.

\section*{Acknowledgement}
This work is financed on the basis of the budget passed by the Saxon State Parliament. The work has also been funded by the German Ministry of Education and Research, joint project: 6G Integrated Communication \& Sensing for Mobility – 6G-ICAS4Mobility, funding label 16KISK231. Furthermore, A. Chorti was supported by INEX funding of excellence.
\bibliographystyle{IEEEtran}

\bibliography{bib}
\end{document}